# The derivation of Particle Monte Carlo methods for plasma modeling from transport equations.

Savino Longo

Dipartimento di Chimica dell'Università and CNR/IMIP, Via Orabona 4, 70126 Bari, Italy.
e-mail: savino.longo@ba.imip.cnr.it

**Abstract**: We analyze here in some detail, the derivation of the Particle and Monte Carlo methods of plasma simulation, such as Particle in Cell (PIC), Monte Carlo (MC) and Particle in Cell / Monte Carlo (PIC/MC) from formal manipulation of transport equations.

Keywords: Charged Particle Transport, Boltzmann Equation, Monte Carlo Methods, Particle-in-Cell, Vlasov/Boltzmann equation

# 1. Introduction

An accurate calculation of the velocity distribution of charged particles in non equilibrium plasmas is necessary in order to evaluate the rate at which collisional elementary processes take place in these media as well as their transport properties [1]. In doing so, one has to take into account the effect of inertia, scattering, externally applied and self-coherent fields, all affecting the particle transport.

To this aim, different particle simulation techniques have been used extensively, as an alternative to the grid-based solution of transport equations [1] by finite differences or finite element techniques. To these belong the very well-know Test Particle Monte Carlo (TPMC, or simply Monte Carlo, MC [1]), the Particle in Cell (PIC [2,3]) and Particle in Cell / Monte Carlo methods (PIC/MC [4]).

No explicit description of the methods will be given here because of their widespread application within the plasma modelling community, and we assume in the following that they are already known to the reader.

Despite the application of these methods is often regarded as a kind of numerical experiment, in view of the convincing 'physical' behaviour of the mathematical particles which are employed, they were derived originally formal manipulations of transport equations, in the field of neutronics for MC, and fusion plasma physics for PIC.

In this paper a discussion of the relation of this particle and stochastic techniques to very well known transport equation is proposed, on the basis of the excellent discussions of [7,15,16] and a recent previous contribution of this same author [11].

In §2 the classical material is reviewed but also critically discussed, in particular for the role of Klimontovich equation and the expression of the collision frequency in MC.

In §3 two different techniques are shown to derive the MC method from the Bolzmann equation in the case of charged particle transport under external fields, also discussing a diagram representations of the related propagator (3a) and the role of the Poisson distribution (3b).

In §4 we extend the treatment to the Vlasov/Boltzmann problem with self-consistent field.

## 2. Preliminaries: PIC method and Monte Carlo

It is already well known that the PIC method provides a solution for the time dependent Vlasov equation [5] in the form:

$$\frac{\partial}{\partial t} f(\mathbf{r}, \mathbf{v}, t) + \mathbf{v} \cdot \nabla_\mathbf{r} f + q(\mathbf{E}/m) \cdot \nabla_\mathbf{v} f = 0,$$
$$\mathbf{E} = -\nabla_\mathbf{r} \varphi$$
$$\varepsilon_0 \nabla^2 \varphi = -\rho \quad (1)$$
$$\rho = q \int d^3 v f(\mathbf{r}, \mathbf{v}, t) + \rho_0$$

Where f is the distribution function and $\rho_0$ is the charge density background due to opposite charged species which are not explicitly simulated.
Namely, the form of the solution considered in this case is the following:

$$f(\mathbf{r}, \mathbf{v}, t) = \frac{w}{N} \sum_{p=1}^{N} \delta(\mathbf{r} - \mathbf{r}_p) \delta(\mathbf{v} - \mathbf{v}_p) \quad (2)$$

This is a solution in terms of N 'quasi-particles' or 'super-particles' each with a 'weight' w. We neglect here complications due to the shape factor of the grid cell [2,4] (our considerations apply equally to any particle/mesh interpolation strategy, and even to a N-body molecular dynamic approach, which is gridless albeit computationally expensive [3]). When such a solution is considered for the Vlasov Equation, this last, in recent times, has been renamed 'Klimontovich equation' after [6] and considered as a starting point to theories for non-ideal plasmas. This distinction can be convenient, but it has the drawback to obscure that the 'Klimontovich equation' is just the Vlasov equation with a special form for its solution, involving generalized functions. Anyway, since the PIC method is able to solve the Klimontovich equation, it can describe accurately any plasma (in the classical regime.
In other terms, the PIC method provides a solution of the Vlasov equation also in the case of a few quasi-particles (even just one, but the solution is trivially analytic up to two particles, obviously): we note here, sideways, that this point has been somewhat neglected in the literature and people often consider the N -
-> infinity limit an essential condition for the validity of (pure) PIC simulations.

This condition, for the pure PIC method, is only necessary when one wants to neglect the micro structure of the charge distribution, and therefore the microfield is to be considered a stray field, which must became negligible with respect to an appropriately averaged electric field. No further discussion of this point is requested for the aim of the present work.

The basic PIC method has been extended in the past (an extensive review can be found in [4]) in order to include the effect of collisions of charged particles with neutral particles, described on the basis of the two following hypothesis:

(1) charged to neutral particle collisions are binary and instantaneous, i.e. the collision time can be neglected with respect to the free-propagation time
(2) (*stoßzahlansatz*:) there is no correlation between the neutral and charged particle distributions and no correlation between neutral particles (no short-range order): therefore the collision probability in a short time step h is given by P(h) = 1-exp(νh) where n is the local charged to neutral particle collision frequency.

The resulting techniques is a merging of the PIC and the Test Particle Monte Carlo method, which had been parallel developed in the field of neutronics [1,7], and later applied to plasmas with a fixed electric field [13], i.e. to the problem of linear transport of particles in a field of scattering centers.

## 3. Monte Carlo solution of the LBE. (a) Green functions and diagrams

The starting point of the argument is the (linear) Monte Carlo method, which had been developed for the problem of neutron transport, and it can be shown to solve the linear Boltzmann equation (LBE ,[1]), in the form:

$$\frac{\partial}{\partial t}f + \mathbf{v}\cdot\nabla_{\mathbf{r}}f = -\nu f(\mathbf{v}) + \int d^3 v' \left( p_{\mathbf{v}'\to\mathbf{v}} f(v') \right), \tag{3}$$

where ν(**r**,**v**,t) is the collision frequency, and p(**r**,t) is the transition frequency in the velocity space. The demonstration of the equivalence between TPMC and LBE was shown [7,16] based on convolution theory: being these studies devoted to neutronics one could neglect external fields (gravity and non uniform magnetic field were not considered, being these studies primarily motivated by nuclear fission device development), and set up a demonstrations based on integral operators with kernels in the closed form of exponentials. The collision frequency, furthermore, was calculated assuming that the target particles are unmoving. The resulting formula for the collision frequency, i.e.

$$\upsilon(\mathbf{r},\mathbf{v}) = n(\mathbf{r})\sigma_{tot}(v)v \qquad (4a)$$

where n is the local target particle number density, has been pupularized for use in MC and PIC/MC simulations by hundreds of papers (for ex.[2,17]) and several textbooks for universal use in cold plasma simulations. The use of this formula actually produces a very strong simplification of the problem, but limits the range of applicability of the method.

Anyway, virtually all the applications of the MC or PIC/MC methods to electron/ion transport in gas plasmas were based on eq.(4a) and sometimes, regrettably, in conditions in which it could not hold, i.e. in which either the target species drift speed (e.g. expanding plasmas) or the static temperature (e.g. ambipolar diffusion of thermal ions) is not negligible.

Namely, the limitations of eq.(5a) were overcome more than 20 years ago by Nanbu [8] and Koura [9] assuming the non linear Boltzmann equation as a starting point: in this case p is a functional of the velocity distribution function of target particle, whose specific expression is

$$p_{\mathbf{v}\to\mathbf{v}'}(\mathbf{r},t) = \int d^3V \int d^3w' T_{\mathbf{v}\mathbf{v}'}^{\mathbf{V}\mathbf{V}'} \delta^4(\mathbf{p}'-\mathbf{p}) F(\mathbf{r},\mathbf{V},t)$$
$$\upsilon(\mathbf{r},\mathbf{v}) = \int d^3v' p_{\mathbf{v}\to\mathbf{v}'}(\mathbf{r},t) \qquad (4b)$$

which includes the matrix element of the T transition operator (which also incorporates all the quantum-statistical normalization factors, and we give the expression for elastic collisions only, for simplicity) [10]. Eq.4b can also be rewritten using the differential cross section:

$$p_{\mathbf{v}\to\mathbf{v}'}(\mathbf{r},t) = \int d^3w\, g\sigma(g,\vartheta) F(\mathbf{r},\mathbf{w})$$
$$\upsilon(\mathbf{r},\mathbf{v}) = \int_0^\pi \sin\vartheta d\vartheta \int d^3w\, g\sigma(g,\vartheta) F(\mathbf{r},\mathbf{w}) \qquad (4c)$$

where **g** is the relative velocity and ϑ is the angle between **g** and **g'**. These last expressions should always be used in place of eq. 4a, when the last cannot be applied.

In the case of charged particle transport one can not neglect the effect of force fields (electric and magnetic) on the particle transport: an appropriately modified methodology based on perturbation theory for the green function has been reported by the present author recently [11].

This demonstration will be reviewed shortly in the following for clarity.

In short, in [11] the present author detached from the Boltzmann Equation a reduced version in which only the negative part of the scattering operator is included, i.e.

$$\frac{\partial}{\partial t} f + \mathbf{v} \cdot \nabla_{\mathbf{r}} f + (\mathbf{F}/m) \cdot \nabla_{\mathbf{v}} f =$$
$$- \upsilon f(v) + \int d^3 v' \left( p_{\mathbf{v}' \to \mathbf{v}} f(v') \right) \quad (5a)$$

$$\frac{\partial}{\partial t} f + \mathbf{v} \cdot \nabla_{\mathbf{r}} f + (\mathbf{F}/m) \cdot \nabla_{\mathbf{v}} f = - \upsilon f(\mathbf{v}) \quad (5b)$$

After calling G the Green function of eq.5a, or *exact* propagator (with a terminology from quantum electrodynamics [18]), and g the Green function of eq.5b, or *inexact* propagator, it is easily found, based on the property of the δ 'function' (but also intuitively clear since the solution propagates along characteristics lines) that g is given by

$$g(\mathbf{r}, \mathbf{r}_0, \mathbf{v}, \mathbf{v}_0, t, t_0) = \delta(\mathbf{r} - \mathbf{r}_p(t)) \delta(\mathbf{v} - \mathbf{v}_p(t))$$
$$\exp\left(-\int_{t_0}^t dt' \upsilon(\mathbf{r}(t'), \mathbf{v}(t'), t')\right) H(t - t_0)$$
$$\frac{dr_p(t)}{dt} = \mathbf{v}_p, \quad m \frac{d\mathbf{v}_p(t)}{dt} = F(\mathbf{r}_p, \mathbf{v}_p, t) \quad (6)$$
$$\mathbf{r}_p(t_0) = \mathbf{r}_0, \quad \mathbf{v}_p(t_0) = \mathbf{v}_0$$

where H(t) is the Heaviside step function, (H(x)=1 if t≥ 0, H(t)=0 if t<0) and can be calculated by tracking a test particle with the appropriate charge to mass q/m ratio starting from the initial point ($\mathbf{r}_0, \mathbf{v}_0, t_0$) under the action of the force field F while calculating the exponential. Furthermore, a straightforward calculation shows that there is a connection between G and g in the form of an integral (Dyson-like [12]) equation

$$G(\mathbf{r},\mathbf{r}_0,\mathbf{v},\mathbf{v}_0,t,t_0) = g(\mathbf{r},\mathbf{r}_0,\mathbf{v},\mathbf{v}_0,t,t_0) +$$
$$\iiint d^3r_1 d^3v_1 d^3v_2 \int_{t_0}^{t} dt_1 g(\mathbf{r},\mathbf{r}_1,\mathbf{v},\mathbf{v}_2,t,t_1) \quad (7)$$
$$p_{\mathbf{v}_1 \to \mathbf{v}_2}(\mathbf{r}_1,t_1) G(\mathbf{r}_1,\mathbf{r}_0,\mathbf{v}_1,\mathbf{v}_0,t_1,t_0)$$

which can be solved by iteration, obtaining the following (Born [12]) series:

$$G(\mathbf{r},\mathbf{r}_0,\mathbf{v},\mathbf{v}_0,t,t_0) = g(\mathbf{r},\mathbf{r}_0,\mathbf{v},\mathbf{v}_0,t,t_0) +$$
$$\iiint d^3r_1 d^3v_1 d^3v_2 \int_{t_0}^{t} dt_1 g(\mathbf{r},\mathbf{r}_1,\mathbf{v},\mathbf{v}_2,t,t_1)$$
$$p_{\mathbf{v}_1 \to \mathbf{v}_2}(\mathbf{r}_1,t_1) g(\mathbf{r}_1,\mathbf{r}_0,\mathbf{v}_1,\mathbf{v}_0,t_1,t_0) +$$
$$+ \iiint d^3r_1 d^3v_1 d^3v_2 \int_{t_0}^{t} dt_1 \iiint d^3r_2 d^3v_3 d^3v_4 \int_{t_0}^{t} dt_2$$
$$g(\mathbf{r},\mathbf{r}_2,\mathbf{v},\mathbf{v}_4,t,t_2) \quad (8)$$
$$p_{\mathbf{v}_1 \to \mathbf{v}_2}(\mathbf{r}_1,t_1) g(\mathbf{r}_1,\mathbf{r}_0,\mathbf{v}_1,\mathbf{v}_0,t_1,t_0)$$
$$p_{\mathbf{v}_3 \to \mathbf{v}_4}(\mathbf{r}_2,t_2) g(\mathbf{r}_2,\mathbf{r}_1,\mathbf{v}_3,\mathbf{v}_2,t_2,t_1) + ...$$

This last equation can be expressed also in the following diagram [12] form, where the bold arrow represent the exact propagator G with the related initial and final points in the (**r**,**v**,t) space, the light arrow represents the inexact propagator g, and the bold dot represents the convolution operation for a single 'p' factor:

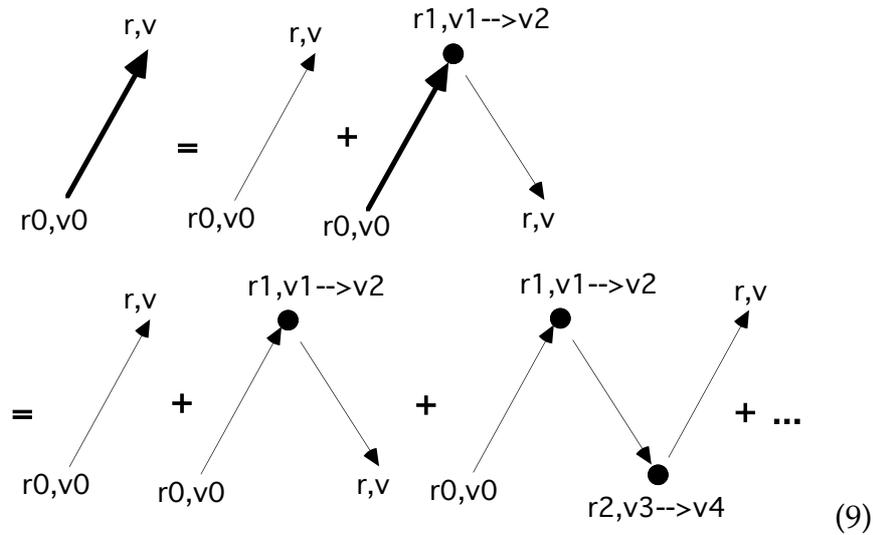

(9)

The Test Particle Monte Carlo method now appear as a technique to calculate the nested integrals implied by the series, using the classical von Neumann method [3,16] which in turn consists in a statistical estimation of the mean value of the integrated function (an alternative clue to close the demonstration is to use a modified version of the 'test function' argument of [1], which is closer to the technique effectively used in applications, but more involved). The diagrams reported above can now be put in conceptual correspondence with the simulated trajectories of test particles in the computation: the number of simulated collisions corresponds to the one of 'dot' operators in the diagram.

**3(b) An alternative derivation showing the 'hidden' role of the Poisson distribution.**

In this section we will derive the MC technique from the transport equation following a different path, which has two peculiarities: (I) it allows to derive directly the so-called 'null collision' MC of Skullerud [13], which is the most frequently used for electron/ion transport simulation, and (II) on a more conceptual framework, it related the hypothesis of 'lack of correlation' (§2) with a Poisson distribution of waiting times, thereby delivering directly the well-known 'time to next collision' formula (eq.18 below) used by most modellers.
To this aim we introduce the majorant transition frequency

$$v_{max} = \max_{\mathbf{r},\mathbf{v}} v(\mathbf{r},\mathbf{v}) \tag{10}$$

and the total derivative:

$$\frac{d}{dt} = \frac{\partial}{\partial t} + \mathbf{v} \cdot \nabla_{\mathbf{r}} + \frac{F(\mathbf{r},\mathbf{v},t)}{m} \cdot \nabla_{\mathbf{v}} \tag{11}$$

Now we can write the eq.(5a) in the form

$$\frac{d}{dt} f(x) = -v_{max} f(x) + v_{max} M f \tag{12}$$

where the action of the 'mixing' operator M on the function f is given by:

$$Mf = \left(1 - \frac{v(\mathbf{r},\mathbf{v})}{v_{max}}\right) f(\mathbf{r},\mathbf{v}) + \frac{1}{v_{max}} \int p_{\mathbf{v}' \to \mathbf{v}} f(\mathbf{r},\mathbf{v}') d^3 v' \tag{13}$$

It is very simple to recognise M as the continuous generalization of a Markov matrix [19]. The (re-)distribution function contained in M has indeed has the appropriate normalization:

$$\int \left[\left(1 - \frac{v(\mathbf{r},\mathbf{v})}{v_{max}}\right) \delta(\mathbf{v} - \mathbf{v}') \delta(\mathbf{r} - \mathbf{r}') + \frac{p_{\mathbf{v} \to \mathbf{v}'}}{v_{max}}\right] d^3 v' = 1 \tag{14}$$

Now we can write the formal solution of the differential equation (12) in the usual form:

$$f(\mathbf{r},\mathbf{v},t) = e^{-v_{max}(1+M)t} f(\mathbf{r},\mathbf{v},0) \tag{15}$$

and, by expanding the exponential in a power series, we find:

$$f(\mathbf{r},\mathbf{v},t) = \sum_{n=0}^{\infty} e^{-v_{max} t} \frac{(v_{max} t)^n}{n!} M^n f(\mathbf{r},\mathbf{v},0) \tag{16}$$

with the position $\mu = v_{max} t$, the final result is:

$$f(\mathbf{r},\mathbf{v},t) = \sum_{n=0}^{\infty} q(n) M^n f(\mathbf{r},\mathbf{v},0)$$

$$q(n) = e^{-\mu} \frac{\mu^n}{n!}$$

(17)

From this last formula, the null-collision Monte Carlo method follows in a remarkably straightforward way, if one considers that the nth power of the matrix M corresponds to the successive application of n 'mixing operation' on the distribution functions as described above. It is in fact shown by eq.(17) that probability, for a simulated particles sampling f(**r**,**v**,t), to be shifted n times to new velocities by the application of the operator M, and therefore according to eq.(13), is given by the Poisson distribution. Such a distribution of events is indeed obtained by generating the succession of collision times {ti} for any simulated particles by the usual formula [11,13,17]

$$t_i = -\nu_{max}^{-1} \ln \zeta_i$$

(18)

where {$\zeta_i$} is a succession of uncorrelated random deviates, uniformly distributed in the interval (0,1). Furthermore, under the application of M, a fraction (1-$\nu(\mathbf{r},\mathbf{v})/\nu_{max}$) of the distribution is left unaltered, coherently with the application of the null-collision method [11,13,17].

We have so derived the null-collision MC technique again and, sideways, we have shown that a linear Boltzmann equation, as well as any Master-like Equation, hides in itself the hypothesis of a Poisson distribution for the occurrence times of 'collision events' which can be rigorously described in terms of a 'mixing operator'.

**4.Application to the Vlasov/Boltzmann problem**

One advantage of the derivations reported in the previous two sections is that the equivalence of the Monte Carlo method to the Linear Boltzmann equation is shown for finite times, and it is therefore sharply different from the heuristic arguments which are usually found in textbooks to justify the application of the numerical method in the course of a short time step h.

One could suspect that the PIC/MC method is actually a solution method for the Vlasov/Boltzmann equation in the form

$$\frac{\partial}{\partial t} f(\mathbf{r},\mathbf{v},t) + \mathbf{v}\cdot\nabla_\mathbf{r} f + q(\mathbf{E}/m)\cdot\nabla_\mathbf{v} f =$$
$$-\upsilon f(\mathbf{v}) + \int d^3 v' \left(p_{\mathbf{v}'\to\mathbf{v}} f(v')\right), \quad (19)$$
$$\mathbf{E} = -\nabla_\mathbf{r}\varphi$$
$$\varepsilon_0 \nabla^2 \varphi = -q\int d^3 v f(\mathbf{r},\mathbf{v},t) + \rho_0$$

Provided a large number of particles is used in the simulation.
Unfortunately, the derivation technique of [1,7,11] cannot be applied directly to the VB equation, because the last is non linear, and therefore the linear perturbation used above cannot be applied.
At the same time, if the number of simulated particles is very high, it is possible to detect a particle subset Z with cardinality N' << N which is at the same time large enough to sample statistically the integrals of eq.(8) (the relative statistical error being in the order of $N'^{-1/2}$ [3]) and small enough to prevent affecting sensibly the space charge and the self-coherent electric field by its evolution. Such a subset can be considered to solve the linearized VB equation written in the form

$$\frac{d}{dt} f_\alpha = -\upsilon f_\alpha(\mathbf{v}) + \int d^3 v' \left(p_{\mathbf{v}'\to\mathbf{v}} f_\alpha(v')\right), \quad (20)$$
$$\varepsilon_0 \nabla^2 \varphi = -q\int d^3 v f_\beta(\mathbf{r},\mathbf{v},t) + \rho_0$$

Where $f_\alpha$ is the contribution of the fraction of the initial charge distribution $f_0$ which is tracked by the α-group of particles under consideration, and $f_\beta = f_0 - f_\alpha$.
The error in confusing the above equation with the original VB equation is due to the neglecting of self-forces on $f_\alpha$ and the effect of $f_\alpha$ on $f_\beta$, and this error will necessarily become negligible when N'/N --> 0.
Since any set of particles large enough can be divided into subsets with the requested properties (not necessarily with an empty intersection, the subset can freely superimpose to one another) we show that the technique of ref. [11] can be applied, and consequently that, under such conditions, the equivalence PIC/MC <-> VB equation holds. This shows our point.

The reader should anyway pay attention, that, on these basis, it is not possible to come back to a Klimontovich-like VB equation, because the δ functions in eq.(2) are changed to ordinary functions under the action of the Boltzmann operator, therefore while the equivalence PIC <-> Vlasov equation holds for any number of particles the extension PIC <--> PIC/MC only holds in the limit N--> inf. Therefore we are leaving here open the problem of detecting the transport equation solved by a few-particle PIC/MC calculation, which, for the moment, can still be considered a 'numerical experiment'.

5.Conclusions

As it was already remarked at the end of ref.[11], Particle/Monte Carlo methods have a double nature, since, while they can be shown, case by case, to provide solutions of known transport equations, they have a so-vivid 'quasi physical' or 'physical analogic' nature that people developing the code are often tempted to consider them as 'numerical experiments' which even allows to 'test' numerical methods based on grid or spectral solutions transport equations. In holding this point of view they have been for a long time in good company in the literature (see or ex. ref. [14]), despite clear expositions such as [1,7] and, ultimately, the spirit itself of the original method [15,16]. In this paper we have reported an unified framework, based on perturbation technique of linear operator, which can be applied to derive such numerical technique directly from transport equations, also in the case of particle under the effect of force fields and even self-consistent force fields. The author hopes that this paper will contribute, together with many others, to make definitely clear that there is no empirical content in a Monte Carlo simulation beyond what can be extracted by solving more explicitly very celebrated transport equations, since they are fully equivalent. At the same time, these stochastic methods remain the most affordable ones to solve multidimensional transport equations, since they are still the better compromise between simplicity and effectiveness in evaluating multi-dimensional integrals like the one appearing in the Born series of §3a.